\documentstyle[12pt,epsfig]{article}

\def\be{\begin{equation}}
\def\ee{\end{equation}}
\def\ba{\begin{eqnarray}}
\def\ea{\end{eqnarray}}
\def\ga{\mathrel{\raise.3ex\hbox{$>$\kern-.75em\lower1ex\hbox{$\sim$}}}}
\def\la{\mathrel{\raise.3ex\hbox{$<$\kern-.75em\lower1ex\hbox{$\sim$}}}}

\newcommand{\bi}[1]{\bibitem{#1}}
\newcommand{\fr}[2]{\frac{#1}{#2}}

\begin{document}

\baselineskip=16pt
\begin{titlepage}

\begin{center}

\vspace{0.5cm}

\large {\bf Scalar-tensor theories with pseudoscalar couplings}
\vspace*{5mm}
\normalsize

{\bf Victor Flambaum$^{(a)}$, Simon Lambert$^{(b)}$}, and {\bf Maxim Pospelov$^{(b,c)}$  }

\smallskip
\medskip

{\it (a) School of Physics,  University of New South Wales, Sydney, NSW 2052, Australia}

{\it (b) Department of Physics and Astronomy,  University of Victoria,
Victoria, BC, V8P 1A1 Canada}

{\it (c) Perimeter Institute for Theoretical Physics, Waterloo,
Ontario N2J 2W9, Canada}

\smallskip
\end{center}
\vskip0.6in

\centerline{\large\bf Abstract}

We consider the scalar-tensor theories of gravity
extended by the pseudoscalar couplings to matter and gauge fields and 
derive constraints on the $CP$-odd combinations of scalar and pseudoscalar couplings
from laboratory spin precession experiments and from the evolution of 
photon polarization over cosmological distances. 
We show the complimentary character of local and 
cosmological constraints, and derive novel bounds on the 
pseudoscalar couplings to photons from the laboratory 
experiments. It is also shown that the more accurate treatment of the 
spin content of nuclei used in the spin precession experiments allows to tighten 
bounds on Lorentz-violating backgrounds coupled to the proton spin.

\vspace*{2mm}

\end{titlepage}

\section{Introduction}
\setcounter{equation}{0}

The discovery of dark energy \cite{SCP} instigated many developments in 
cosmology and particle physics during the last decade.
To date  all observational data are consistent with the most economic
possibility: the dark energy is just a cosmological constant, and as such does not evolve 
over the cosmological time scales. On the other hand, it is intriguing to think about the 
alternative explanations related to a drastic change of the infrared physics.
In parallel to the attempts of modifying gravity on large scales \cite{LSMG}, 
there is a renewed interest in the cosmological scalar fields that are 
nearly massless, and manifest themselves as a "dark energy" component 
over large cosmological distances \cite{quint}.

An interesting twist to the well-known story of cosmological scalars comes
from the possibility of their interaction with matter and gauge fields
(For purely cosmological signatures of "interacting" quintessence, see {\em e.g.} \cite{Amendola}). 
In fact such theories exhibit 
a rich plethora of phenomena that go beyond pure cosmological effects, 
which we would like to illustrate on the following 
toy example. Let us consider a  Lagrangian for the scalar field
$\phi$  interacting with a 
Standard Model fermion $\psi$ ({\em e.g.} electron) 
and a gauge field $A_\mu$ ({\em e.g.} photon),
\begin{eqnarray}
\label{toy}
 {\cal L} = \frac12\partial_\mu \phi \partial^\mu \phi - V(\phi) + 
   \bar \psi (iD_\mu\gamma^\mu - m)\psi -\fr14 F_{\mu\nu}F^{\mu\nu}\\
  -c_{S\psi }\phi\bar\psi \psi -c_{P\psi } \phi\bar\psi i\gamma_5 \psi
-c_{ S \gamma}\phi F_{\mu\nu}F^{\mu\nu}-c_{P\gamma }\phi  F_{\mu\nu}\tilde F^{\mu\nu}.\nonumber
  \end{eqnarray}
  Here $c_{Si}$ and $c_{Pi}$ parametrize the strengths of the scalar and pseudoscalar couplings
to photons and fermions, while $F_{\mu\nu}$ and $\tilde F_{\mu\nu}$ denote the usual 
and dual field strengths, and $D_\mu$ is the covariant derivative.
  Written in flat space, Lagrangian (\ref{toy}) can be trivially generalized 
  to curved backgrounds, and to nonlinear couplings to matter. 
Starting from (\ref{toy}), one can immediately infer a number of interesting consequences,
a partial list of which is given below.
\begin{enumerate}

\item {\em The existence of a new long-range force distinguishable from spin-two gravity.}
The scalar field contributes to the gravitational force, adding
$\sim c_s^{2}$ on top of the familiar Newton constant mediated by gravitons.
Such a force leaves distinguishable imprints via relativistic corrections and/or composition dependence (effective violation of the equivalence principle).

\item {\em The existence of a preferred Lorentz frame associated with $\partial_t \phi$. } 
If $\phi$ is a very light quintessence-like field, then there is a 
preferred frame where cosmologically $\partial_\mu \phi = (\dot \phi,0,0,0)$.
For most of the models this frame
coincides with the frame of the cosmic microwave background (CMB), 
and $|\dot \phi|$ is limited by $(\rho_{d.e.}(1+w))^{1/2}$, where 
$w$ is the dark energy equation of state parameter. 

\item {\em Variation of masses and couplings in time and space.}
Effective values of masses and coupling constants 
vary in space and time, $m_{phys}(t,{\bf x}) = m+c_{S\psi }\phi(t,{\bf x})$, 
following the $\phi$-profile. 

\item {\em Coupling of polarization to velocity relative to the CMB frame.}
A particle moving relative to the CMB frame  
acquires a helicity-dependent interaction, $ H_{int} \sim ({\bf Sn}) \dot\phi$,
where $n$ is the direction of propagation. This way, the $C_{P\gamma}$-proportional interaction
would result in the rotation of polarization for photons 
propagating over varying $\phi$-background.

\item {\em Photon-scalar conversion.} In the presence of an external electromagnetic field 
a photon  can "oscillate" to a quantum of the scalar field thereby {\em e.g.} reducing
the luminosity of distant objects or providing additional channels for star cooling. 

\item {\em Coupling of spin to the local gravitational force.}
Scalar coupling $g_{S\psi}$ will lead to the local field gradient 
$\nabla \phi$ generated by massive bodies, which is closely parallel to 
the vector of local free-fall acceleration. The pseudoscalar couplings 
then create a Zeeman-like splitting for the spin of $\psi$-particles 
in the direction of the local gravitational 
acceleration, $H_{int} \sim ({\bf Sg})$. 

\end{enumerate}

It is remarkable that such a simple Lagrangian leads to a number of quite different 
phenomena. 
Unfortunately, at this stage the exciting phenomenology of "interacting dark energy" 
lives in a pure theoretical realm:
 there is no confirmed experimental evidence for 
any of the effects on our list\footnote{A tentalizing hint on the redshift evolution of 
the fine structure constant was reported in Ref. \cite{Webb}, 
which so far has not been corroborated by other searches \cite{Petitjean}. Also, an earlier 
claim of the 
non-zero pseudoscalar-induced anisotropy in polarization signal \cite{Ralston} was 
disputed in the literature \cite{AntiRalston}.}. Consequently, there are 
only upper limits on the combinations 
of the couplings in Lagrangian (\ref{toy}) that can be quoted. 
Nevertheless, many of the effects on our list have found an extensive coverage in the theoretical 
works.  Most notably, the changing couplings were 
discussed, for example, in Refs. \cite{ch_alpha,OP,Uzan}, the photon-scalar conversion considered in 
Refs. \cite{ph-sc}, and the fixed frame effects versus the cosmological evolution of photon 
polarization were addressed in a series of papers
\cite{Opt_act,Carroll,Opt_act_cmb,Ralston,AntiRalston}. For the limits on 
scalar-induced corrections to gravitational 
interactions we refer the reader to recent reviews \cite{Will}
and references therein. In contrast, the last item on our list, 
the spin coupling to the local 
gradient of the scalar field received far less attention (see {\em e.g.} \cite{MW}). 

The purpose of this paper is to show that the pseudoscalar couplings of the 
Brans-Dicke type scalar can indeed 
be subjected to stringent laboratory constraints that are complementary to the cosmological 
limits. The high-precision spin precession experiments constrain pseudoscalar 
interactions both in the fermion and photon sectors.
In the rest of this paper we present the set-up for our model, briefly 
review the effects created by the cosmological evolution of $\phi(t)$, investigate
the local spin effects created by the gradient of  $\phi$, and set
 the limits on the admissible size of the pseudoscalar couplings.

Before we delve into studying the physical effects induced by the pseudoscalar couplings, 
we would like to add a word of caution addressed to all models of "interacting 
quintessence". The models of light scalar fields represent a formidable challenge at the 
quantum level, as there are no fundamental reasons 
for a scalar to remain massless or nearly massless. 
The scalar interaction of such field 
makes the whole problem even more difficult, if not impossible, 
from the point of view of "technical naturalness": the loops of  
Standard Model (SM) fields tend to generate big corrections to $V(\phi)$ even with 
a relatively small ultraviolet cutoff parameter, which 
would be in conflict with requirements $m_\phi \sim H$ \cite{OP,Carroll,BDD}. 
There is no clear resolution to this problem, 
which essentially prevents the fully consistent 
study of $\phi$ dynamics. Instead, one has to rely, perhaps too optimistically, 
that the problem of 
near-masslessness of the scalar field could be cured by the same mechanisms 
that make the cosmological constant small and meanwhile keep $V(\phi)$ fixed by hand. 
To finish this "disclaimer" on an optimistic note, we would like to remark that the pseudoscalar couplings do 
not make this problem worse. Indeed, in essence the pseudoscalar couplings give
only derivative interactions, and therefore do
not affect the potential $V(\phi)$ at perturbative level. 


\section{Adding spin couplings to scalar-tensor theories}
We would like to formulate our reference Lagrangian at the normalization 
scale just below the QCD scale, so that the effective matter degrees 
of freedom are electrons, photons, nucleons and neutrinos. 
Splitting the $\phi$-field Lagrangian into the scalar and pseudoscalar parts,
\be
{\cal L} = {\cal L}_S+{\cal L}_P,
\ee
we choose the following parametrization,
\be
{\cal L}_S = \frac12\partial_\mu \phi \partial^\mu \phi - V(\phi) -
\sum_{j=e,p,n}\fr{\phi}{M_{Sj}}m_j\bar\psi_j \psi_j - \fr{\phi}{M_{S\gamma}}F_{\mu\nu}F^{\mu\nu},
\label{scalar}
\ee
and
\be
{\cal L}_P = \sum_{j=e,p,n,\nu}\fr{\partial_\mu\phi}{M_{Pj}}~\bar\psi_j \gamma_\mu\gamma_5\psi_j 
- \fr{\phi}{M_{P\gamma}}F_{\mu\nu}\tilde F^{\mu\nu}.
\label{pseudo}
\ee
Lagrangian (\ref{pseudo}) includes all possible pseudoscalar interaction at mass 
dimension five level. 
Notice that the pseudoscalar interactions can be chosen in a slightly
different form, $\bar\psi \gamma_5 \psi$, as in (\ref{toy}). This does not mean, however,
that the set of our operators should be enlarged. The two type of operators,
pseudoscalar and axial vector, are related on the equations of motion. 
These equations are in general anomalous, but since we include the interaction 
with $F\tilde F$ explicitly, we can assert that Lagrangian (\ref{pseudo}) 
is indeed complete in a given dimension of the operators. 

The scalar part of the Lagrangian (\ref{scalar}) leads to new contribution to gravitational 
force, and to change of masses and couplings. Since in this paper our main interest is in 
spin effects, we are going to make simplifying assumptions of approximate universality of the 
$\phi$-mediated attractive force,
\be
M_{Se}=M_{Sp}=M_{Sn}\equiv M_S, ~~~{\rm and}~~~ M_{S\gamma}\gg M_S.
\ee
At distances shorter than the Compton wavelength of $\phi$-quanta the 
Newton constant receives contributions from both spin-two and spin-zero exchanges, 
\be
G_N = G_N^0\left(1+ \fr{2M_{\rm Pl}^2}{M_S^2}\right),
\ee
where $G_N^0$ is the unperturbed Newton constant due to graviton exchange, and 
Planck mass is defined as $M_{\rm Pl} = (8\pi G_N^0)^{-1/2}=2.4\times 10^{18}$GeV. 

If needed, the pseudoscalar couplings could be "lifted" from the nucleon level to the 
level of individual quarks. Using the experimental results for the spin content of the 
nucleon combined with $SU(3)$-flavour relations \cite{Ellis} one gets
\be
M_{Pp}^{-1} \simeq  0.8 M_{Pu}^{-1} - 0.4 M_{Pd}^{-1} - 0.1 M_{Ps}^{-1} ;~~~ M_{Pn}^{-1} \simeq  
0.8 M_{Pd}^{-1} - 0.4M_{Pu}^{-1}  - 0.1 M_{Ps}^{-1},
\ee
where the light quark couplings are normalized at the scale of 1 GeV. 

Using the appropriate field content, one can determine the renormalization group 
evolution of the pseudoscalar couplings. In general the equations
governing this evolution  takes the 
following form,
\begin{eqnarray}
\nonumber
\frac{dM^{-1}_{Pi}}{d\log(\Lambda/\mu)} &=& a_{ij} M^{-1}_{Pj} + b_{i\alpha} M^{-1}_{P\alpha} \\
\frac{dM^{-1}_{P\alpha}}{d\log(\Lambda/\mu)} &=& c_{\alpha i}M^{-1}_{Pi}  + d_{\alpha\beta} M^{-1}_{P\beta},
\label{renorm}
\end{eqnarray}
where the logarithm is taken between the ultraviolet scale $\Lambda$ and the infrared scale $\mu$,
Latin indices indicate fermionic fields and Greek indices indicate the gauge bosons 
of the SM group. The renormalization group coefficients $a_{ij}$, $b_{i\alpha}$, $c_{\alpha i}$,
and $d_{\alpha\beta}$ depend on charge assignments and coupling constants of field running 
inside the loops. The precise form of these coefficients is not of immediate interesting 
to us, but we would like to emphasize the following important observation:
at any loop level the derivative couplings to fermions {\em do not} generate 
couplings to $F_{\mu\nu}\tilde F^{\mu\nu}$. In other words,
\be
c_{\alpha i} \equiv 0.
\ee
Whatever size of the pseudoscalar couplings between photons and $\phi$ 
is generated by some (perhaps anomalous)
ultraviolet scale physics at energies order $\Lambda$, it is preserved by the subsequent 
evolution to the lower scales. In fact, 
this refers both to the logarithmic running and to the threshold corrections. 
This observation delineates two important classes of models: there are models 
where both fermion and photon pseudoscalar couplings present in the Lagrangian, 
and there are models where only couplings to fermions are present. The models 
where $\phi$ couples only to gauge bosons would necessarily be fine-tuned, as 
quantum effects in (\ref{renorm}) would definitely generate induced couplings to fermions. 

Existing constraints on the model can be devided into pseudoscalar and scalar constraints. 
The constraints on the universal scalar coupling $M_S$ can be 
derived from the constraint imposed by the Cassini satellite data on the post-Newtonian
parameter $\bar\gamma$ \cite{Cassini},
\be
|\bar\gamma|<4\times 10^{-5} ~~~\Longrightarrow~~~ M_S > 400~M_{\rm Pl}.
\label{Cassini}
\ee
The constraints on the non-universal part of the scalar coupling are several 
orders of magnitude stronger. 
The scalar coupling to photons is constrained via the limits on the time variation 
of the coupling constant, and less directly via the composition-dependent 
contribution to local accleration. Typically, one has $M_{S\gamma} > 10^3 M_{\rm Pl}$. 
In contrast, the pseudoscalar couplings are far less constrained. The leading source of constraints
are the energy losses mechanisms in stars \cite{Raffelt}, and for electrons, photons and 
nucleons all constraints are in the ballpark of 
\be
\label{limstars}
|M_P| \ga (10^{10}-10^{12})~ {\rm GeV} \sim (10^{-8}-10^{-6})\times M_{\rm Pl}.
\ee 
In the next section, we are going to show that if both pseudoscalar and scalar couplings 
are present,  some constraints on $M_P$ can be significantly improved.

\section{Cosmological constraints on the model}
\setcounter{equation}{0}


To derive cosmological constraints on pseudoscalar couplings we 
remind the reader that the presence of a time-evoloving scalar 
field with a pseudoscalar coupling to photons leads 
to a rotation of polarization for photons. The resulting angular
change in the linear polarization for a photon propagating from 
point 1 to point 2 is simply related to the change of $\phi$
between the two points,
\be
\Delta \theta = \frac{2\Delta \phi}{M_{P\gamma}}.
\ee
Following the work of Carroll \cite{Carroll} and the original analysis
of Ref. \cite{Kronberg}, we use the limit on the extra rotation of 
polarization from distant source (3C 9) at redshift $z=2.012$
as $|\Delta \theta|<6^\circ$,
\be
\fr{|\phi(z=2)-\phi(z=0)|}{M_{P\gamma}} < 0.052.
\label{low_z_limit}
\ee

Even more distant sources of polarization are available in the studies of the
cosmic microwave background. The $E$-mode polarization map of the sky has 
been produced \cite{WMAP-pol}, which agrees well with the expectation based on the temperature 
map. This constrains the amount of extra rotation of polarization 
introduced by $\phi F\tilde F$ interaction between the surface of last 
scattering and $z=0$. Recent numerical analyses of the CMB data 
provide a constraint on the amount of extra rotation at the level of 
 $|\Delta \theta|<6^\circ$ \cite{Opt_act_cmb} (the same limit of $6^\circ$ is purely coincidental),
which allows to extend (\ref{low_z_limit})
 to the redshifts of photon decoupling, $z_{dec}\simeq 1100$,
 \be
\fr{|\phi(z_{dec})-\phi(z=0)|}{M_{P\gamma}} < 0.052.
\label{high_z_limit}
\ee

Finally, we would like to point out that the CMB polarization signal 
is generated in the narrow window of redshifts that correspond to the "last scattering" 
surface, and therefore the existing measurements constrain the amount of extra rotation 
within the thickness of this surface,
\be
\frac{2}{M_{P\gamma}} \left|\Delta\phi(z_{dec}\pm\Delta z_{dec}/2)\right|<O(1),
\label{thickness}
\ee
where $\Delta z_{dec}\simeq 200$ correspond to the thickness of the last scattering
surface. The violation of this bound would {\em suppress} the strength of polarization signal,
which is well measured. 

With these bounds at hand, we are ready to translate them into the constraints on the 
parameters of our model. However, the cosmological constraints 
depend very sensitively on what we assume 
about the scalar couplings of $\phi$ to dark matter and even more so on the 
choice of the potential $V(\phi)$. Since the number of options is infinite, 
we would like to consider in detail two well-motivated cases.\\
{\em Case 1.} The simplest case is when the potential for $\phi$ is nearly flat and
the evolution of $\phi$ is slow. In this case one can linearize $V(\phi)$,
\be
V(\phi) \simeq \rho_\Lambda\left(1 +\fr{\phi}{M_\Lambda}\right),
\ee
where $\rho_\Lambda$ is approximately equal to the measured value of 
dark energy density, and $M_\Lambda$ is a new parameter on the order of the 
Planck scale and/or $M_S$. 
In the limit when the  back-reaction of $\rho_\phi$ on Friedmann's equations is neglected 
one can find an analytic expression for the evolution of $\phi$ in the flat Universe \cite{OP}. 
In this approximation the time evolution 
of the scale factor can be expressed via the scale factor and the
Hubble parameter today ($t_{now}\equiv t_0$): $H_0=H(t=t_0)=\dot a/a|_{t=t_0}$  and $a(t=t_0)\equiv a_0$,
as well as the current energy densities of matter and cosmological constant relative to the 
critical density,  $\Omega_m=\rho_m/\rho_c$ and $\Omega_\Lambda=\rho_\Lambda/\rho_c$:
\begin{equation}\label{a}
a(t)^3=a_0^3\frac{\Omega_m}{\Omega_\Lambda}\left[\sinh(\frac{3}{2}\Omega_\Lambda^{1/2}H_0t)\right]^2.
\end{equation}
The equation of motion for the scalar field receives forcing terms directly related to dark energy 
and matter densities:
\begin{equation}\label{cosmo}
\ddot{\phi} +3H\dot{\phi} = - \frac{\rho_m}{M_S} -\fr{\rho_\Lambda}{M_\Lambda} 
= -\rho_c\left[\fr{\Omega_m}{M_S}\left(\frac{a_0}{a}\right)^3+\fr{\Omega_\Lambda}{M_\Lambda}\right],
\end{equation}
where we made an assumption of the universal strength of $\phi$ coupling to matter, 
including the dark matter. 
This equation can be integrated out explicitly \cite{OP} to give 
\begin{equation}\label{phi}
\phi(t)=\frac{4}{3}M_{\rm Pl}^2\left[\left(\frac{1}{2M_\Lambda}-\fr{1}{M_S}\right)(bt_0 \coth(bt_0)-bt \coth (bt)) -
\fr{1}{M_S} \ln \frac{\sinh(bt)}{\sinh(bt_0)}  \right],
\end{equation}
where the following notation has been introduced:
\begin{eqnarray}
b = \frac{3}{2} \Omega_\Lambda^{\frac{1}{2}}H_0.
\end{eqnarray}
This solution implies boundary conditions $\dot\phi|_{t\rightarrow 0}\sim$ not too large and $\phi(t_0)=0$. The first condition 
is automatically satisfied as $\phi$ does not evolve rapidly during the radiation domination, and the second condition
is simply a choice, possible since $\phi$ enters linearly in the Lagrangian. It is easy to see that in the 
limit of $t\ll t_0$ the dependence of $\phi$ on  redshift is logarithmic,
\be
\phi(a) \simeq {\rm const} - \fr{2M_{\rm Pl}^2}{M_S}\ln(a/a_0)~~~{\rm at }~~~t_{eq}\ll t\ll t_0.
\label{early_t}
\ee

Now we can use the evolution (\ref{phi}) to impose limits on the combination of $M_P$ and $M_{S(\Lambda)}$
parameters using observational constraint (\ref{low_z_limit}). 
We do it for three separate representative cases: for the equal couplings to the matter and 
dark energy density, and for couplings to dark energy and matter only:
\begin{eqnarray}
\label{new1}
M_\Lambda=M_S~~~~~~~~~~~~ |M_{P\gamma}M_S|&>&36~M_{\rm Pl}^2;\\ 
\label{new2}
|M_\Lambda|\to\infty~~~~~~~~~~~~ |M_{P\gamma}M_S|&>&30~M_{\rm Pl}^2;\\ 
|M_S|\to \infty~~~~~~~~~~~~ |M_{P\gamma}M_\Lambda| &>&6.1~M_{\rm Pl}^2. 
\label{old}
\end{eqnarray}
These limits generalize the analysis of Ref. \cite{Carroll} where only the $V(\phi)$-induced 
optical rotation was considered. We also notice that both (\ref{new1}) and (\ref{new2}) are about one 
order of magnitude stronger than (\ref{old}), which is the consequence of $(a_0/a)^{3}\sim 8$ enhancement of
matter density over the cosmological constant at redshifts $\sim 2$. 

Due to logarithmic dependence 
on redshifts at $t\ll t_0$ (\ref{early_t}) there is about one order of magnitude gain
in the strength of the constraint when using the CMB limit (\ref{high_z_limit})
for the case of finite $M_S$,
\be
 M_\Lambda=M_S ~~~{\rm or}~~~ |M_\Lambda|\to\infty ~~~\Longrightarrow ~~~|M_{P\gamma}M_S|~>~255~M_{\rm Pl}^2. 
\label{new3} 
\ee

In order to see the maximal sensitivity to $M_{P}$, we can saturate the constraint on $M_S$ 
(\ref{Cassini}), which results in 
$M_{P\gamma} \ga O(M_{\rm Pl})$ at maximally allowed $M_S$. 

{\em Case 2.} Going away from the linearized case,  
 we consider the cosmological evolution of $\phi$-field approaching some
local minimum of $V(\phi)$, 
\be
V(\phi) = \rho_\Lambda + \fr{m_\phi^2}{2}(\phi-\phi_0)^2.
\ee
If the mass of the field is well above the current Hubble parameter, $m_\phi \gg H_0$, then 
the evolution of $\phi$ starts long before the present epoch.
A well-known solution for $\phi$ in this case are the oscillations around
the minimum with the amplitude that red-shifts as $a^{-3/2}$. If the initial deviation of $\phi$ from 
equilibrium was $\phi_{in}$ at the time $t_{in}$ when oscillations began, $H(t_{in})\sim m_\phi$,
then the subsequent evolution in the radiation domination will be given by
\be
\phi(t)\simeq \phi_0 + \phi_{in} \left( \fr{a_{in}}{a(t)}\right)^{3/2} 
\!\!\!\!\cos[m_\phi(t-t_{in}) + \alpha],
\label{osc}
\ee
where $\alpha$ is some phase factor. Because of the red-shifted amplitude in (\ref{osc}), 
the constraints provided by the CMB are clearly more advantageous than the low $z$ constraints.
However, the oscillations of $\phi$ (\ref{osc}) make it difficult to define $\phi(z_{dec})$, and consequently
the analyses of \cite{Opt_act_cmb} with limits (\ref{high_z_limit}) are not directly applicable
and instead one should resort to limits (\ref{thickness}). Still, if the initial deviation of 
$\phi$ field from its minimum is on the order or less than the pseudoscalar coupling $M_{P\gamma}$,
and oscillations begin earlier than the decoupling, then the cosmological evolution of polarization 
provides {\em no constraints} on the size of the pseudoscalar coupling, 
\be
|\phi_{in}|<|M_{P\gamma}|;~~~ t_{in} \ll t_{dec}~~~\Longrightarrow ~~~ {\rm no~constraints~on}~~M_{P\gamma}.
\label{noconstr}
\ee
This is an important observation, since the first condition $|\phi_{in}|<|M_{P\gamma}|$ is quite natural 
if $\phi$ field has a phase-like origin similar to {\em e.g.} QCD axion, and $t_{in} \ll t_{dec}$ is 
satisfied for all masses of $\phi$ in excess of $10^{-28}$ eV.

\section{ Local spin precession constraints}

As we have shown in the two previous sections, the cosmological constraints 
on pseudoscalar couplings apply only to $M_{P\gamma}$, and not to fermionic couplings. 
Moreover, all cosmological constraints will be eliminated if the field starts oscillating 
much earlier than the decoupling of the CMB photons (\ref{noconstr}). 
This leaves a large domain of parameter 
space where only the local experiments are 
going to be sensitive to the pseudoscalar couplings. 
We wish to consider them in this section. Before we do that, we would like to note that the 
couplings of spins to the local gravitational (spin-two) field 
has been extensively studied in the literature
\cite{sg-old,sg-med,sg-new}. Of main interest for us is the conclusion 
reached in these works that ${\bf g\cdot S}$ coupling does not arise in general relativity.
Therefore, if detected, it can be thought of as a distinct signature of the scalar exchange. 

Since most of the experiments deal with non-relativistic atoms and nuclei, it is 
convenient to use the non-relativistic Hamiltonian, 
\be
H_{int} = - \sum_{j=n,p,e} \fr{ (\sigma_j\cdot \nabla\phi)}{M_{Pj}} 
+\int d^3x\fr{4({\bf E\cdot B})\phi}{M_{P\gamma}},
\label{H_int}
\ee
where $\vec{\sigma} = {\bf S}/|S|=2 {\bf S}$. 
The local gradient of $\phi$ is one-to-one related to the gravitational acceleration,
\be
\nabla \phi ={\bf g}~ \fr{2M_{\rm Pl}^2}{M_S},
\ee
so that the strength of interaction of each spin to the gravitational 
field is given by $g \times 2M_{\rm Pl}^2/(M_SM_{Pj})$. Gravitational 
acceleration has dimension of energy in particle physics units of $c=\hbar=1$, 
and corresponds to the frequency splitting of spin up and spin down states 
$\nu_g =2\times  9.8 ~10^2$cm/s$^2/(2\pi\times 3~10^{10}$cm/s$) = $10.4 nHz. 
Unlike most  problems in quantum mechanics where "up" and "down" are 
usually a matter of convention, in this theory 
these words should be used literally. 
Only a handful of spin precession experiments ever reached the sensitivity 
lower than 10 nHz, among them the experiments searching for the permanent electric 
dipole moments of diamagnetic atoms \cite{HgEDM}, where the statistical sensitivity is 
comparable or better than 10 nHz. Unfortunately, this sensitivity is related to the 
energy difference of spins in parallel and anti-parallel electric and magnetic 
fields and does not translate  into the limits on spin interaction with the vertical 
direction. 

Dedicated search for ${\bf g\cdot S}$ interaction was
pursued in \cite{Hg-Hg} (and earlier in \cite{gs}), where a $\sim \mu$Hz accuracy was achieved. 
 In particular, experiment \cite{Hg-Hg} compared the 
precession frequencies of two mercury isotope spins, $^{199}$Hg and $^{201}$Hg for 
different orientations of magnetic field and set a limit of 2.2$\times 10^{-30}$ GeV 
for the spin-dependent component of gravitational energy. 
Another group of measurements that can be 
used to limit the pseudoscalar couplings are the spin precession experiments 
that searched for the effects of Lorentz violation \cite{Berglund,Bear} and the experiment with spin-polarized 
pendulum \cite{Heckel}. The absence of sidereal 
modulation of spin precession, confirmed by these experiments, sets the limit on the coupling 
of spins to any direction in space that {\em does not} change as the Earth rotates around its axis. 
Besides useful limits on Lorentz-violating theories \cite{Kost}, such effects will constrain the pseudoscalar 
couplings in combination with $\nabla \phi$ created by astronomical bodies other than the Earth.
The solar contribution to $\nabla \phi$ is smaller than $\nabla \phi_{Earth}$ by a factor of 
$\sim 6\times 10^{-4}$, 
thereby reducing the strength of the constraints extracted from sidereal 
variations by the same amount. Putting 
different results together, and assuming that the 
range of the force is comparable to or larger than the solar system, 
we arrive at the following set of constraints,
\begin{eqnarray}
\label{constr1}
|M_{Pn}M_S|&>&5\times 10^{-4}~M_{\rm Pl}^2~~~{\rm Ref.}~[29]\\ 
\label{constr2}
|M_{Pn}M_S|&>&1\times 10^{-4}~M_{\rm Pl}^2~~~{\rm Ref.}~[31,32]\\
\label{constr3}
|M_{Pe}M_S|&>&2\times 10^{-5}~M_{\rm Pl}^2~~~{\rm Ref.}~[33]
\end{eqnarray}
Bounds (\ref{constr1}) and (\ref{constr2}) are derived in the assumption of Ref. \cite{KostLane} 
that the spin of the nucleus is given by the angular momentum 
of the of the outside nucleon, which happens to be a neutron for 
all nuclei used in the most sensitive searches ($^3$He, $^{129}$Xe, 
$^{199}$Hg, $^{201}$Hg). Consequently, the limits are formulated on the 
pseudoscalar coupling to neutrons, as it is also the case for the limits on the 
external Lorentz-violating axial-vector backgrounds \cite{KostLane}. 

In fact, one can refine these bounds and impose separate constraints on the strength of the pseudoscalar
coupling for protons and neutrons. Although most of the nuclei in atoms used in experiments 
\cite{Hg-Hg}-\cite{Bear} have a valence neutron outside of closed shells, one can use the information on 
the magnetic moments of these nuclei together with simple theoretical model of nuclear structure
to deduce the proton contribution to the total nuclear spin. To be specific we shall assume that 
 the magnetic moment of the nucleus is composed entirely from 
the spin magnetic moment of the valence neutron and spin magnetism of polarized nuclear core,
\begin{eqnarray}
\label{simplespin}
\mu = \mu_n \langle \sigma_{z}^{(n)} \rangle + \mu_p \langle \sigma_{z}^{(p)} \rangle\\
\langle \sigma_{z}^{(n)} \rangle + \langle \sigma_{z}^{(p)} \rangle = \langle \sigma_{z}^{(0)} \rangle. \nonumber
\end{eqnarray}
In these equations, $\mu,~ \mu_p, ~\mu_n$ are the magentic moment of the nucleus, proton and neutron. 
Numerical estimates show that the orbital
contribution to the  magnetic moment $\mu$ in the nuclei of
interest is less important than the spin contribution since the neutron
orbital contribution is zero and the proton orbital contribution is small
in comparison with the proton spin contribution. The latter is enhanced by the large
value of the proton magnetic moment $\mu_p=2.8$, which justifies the neglection of
proton orbital magnetism for low $l$ obitals.
Neglection of spin-orbit interaction makes total spin conserved and its total 
value equal to the average spin 
of the neutron above the unpolarized core, $\langle \sigma_{z}^{(0)} \rangle$. The latter is 
equal to $1$ for $j=l+1/2$ and $-j/(j+1)$ for $j = l-1/2$, where $j$ is the value of the nuclear angular 
momentum, and $l$ is the orbital quantum number of the valence neutron.  
Using these simple formulae (\ref{simplespin}), we determine  $\langle \sigma_{z}^{(n)} \rangle$
and $\langle \sigma_{z}^{(p)} \rangle$ for observationally relevant cases of 
$^{199}$Hg, $^{201}$Hg, $^{129}$Xe, and $^{3}$He as shown in Table 1.

\begin{table}
\begin{center}
\begin{tabular}{c|c|c|c|c|c}
nucleus & $\mu$ & $j, ~l$ & $\langle \sigma_{z}^{(0)} \rangle$ &
 $\langle \sigma_{z}^{(n)} \rangle$ &  $\langle \sigma_{z}^{(p)} \rangle$\\
\hline\hline
 $^{3}$He   & -2.13 & 1/2, 0 & 1   & 1.04  & -0.04\\
 $^{129}$Xe & -0.78 & 1/2, 0 & 1   & 0.76  & 0.24\\
 $^{199}$Hg & 0.50  & 1/2, 1 & -1/3& -0.31 & -0.03\\
 $^{201}$Hg &-0.56  & 3/2, 1 & 1   & 0.71  & 0.29\\
\hline
\end{tabular}
\caption{\footnotesize Composition of the nuclear spin }
\end{center}
\end{table}

One can see that the constribution of the proton spin into the total 
spin of these nuclei, especially $^{129}$Xe and  $^{201}$Hg, can be as high 
as 30\%, and therefore the proton pseudoscalar coupling is also limited 
in these experiments. For example,  experiment \cite{Hg-Hg}
limits the following combination of the proton and neutron couplings:
\be
\label{prlimit}
|M_{P\rm eff}M_S|>1.5\times 10^{-4}~M_{\rm Pl}^2, ~~{\rm where}~ M^{-1}_{P\rm eff} 
= -0.5 M_{Pn}^{-1} +0.7 M_{Pp}^{-1}.
\ee 
The relative enhancement of the proton contribution is due to a rather close cancellation 
of neutron contribution to the differential frequency of spin precession for $^{199}$Hg and  $^{201}$Hg.

As a bi-product of our analysis,
we can improve the bounds on the Lorentz-violating axial-vector couplings in the
Colladay-Kostelecky parametrization \cite{Kost}. Indeed, the spatial 
components of the axial vector 
background to protons, $b_\mu$, is constrained in the same experiments, Refs. \cite{Berglund,Bear}, 
in particular because of the substantial contribution of proton spin to the spin of 
$^{129}$Xe. For example, the interpretation of the null result of the most sensitive experiment 
\cite{Bear} with the use of the analysis \cite{KostLane} that assumes $\langle \sigma_{z}^{(n)} \rangle=
\langle \sigma_{z}^{(0)} \rangle$, $\langle \sigma_{z}^{(p)} \rangle$, and our 
work differ in the following way:
\begin{eqnarray}
\nonumber
2\pi \nu_{LV} = 2b^{(n)}_i\left(1 - \fr{\mu_{\rm He}}{\mu_{\rm Xe}}\right) = -3.5  b^{(n)}_i, ~~ {\rm Ref. [35]}\\
2\pi \nu_{LV} = 2(0.76b^{(n)}_i +0.24b^{(p)}_i - \fr{\mu_{\rm He}}{\mu_{\rm Xe}}(1.04 b^{(n)}_i-0.04b^{(p)}_i))
= -4.2  b^{(n)}_i + 0.7 b^{(p)}_i~{\rm this~work},
\nonumber
\end{eqnarray}
where $\nu_{LV}= 53\pm 45$ nHz is the experimentally measured 
(and consistent with zero) Lorentz-violating frequency shift \cite{Bear}. 
Obviously, the contribution of $b^{(p)}_i$ to $\nu_{LV}$ is non-negligible, 
and implies that $|b^{(p)}_i| < {\rm few} \times O(10^{-31})$ GeV, which is 
far better than the results of dedicated searches of Lorentz violation in the
proton sector  with {\em e.g.} hydrogen maser \cite{Walsworth}.

Besides the constraints on nucleon and electron couplings, the same clock comparison experiments 
allow to set limits on $M_{P\gamma}$. For example, for an atom (or nucleus) with the 
total angular momentum $J$, the matrix element of $\phi F\tilde F$ interaction 
is not zero,
\be
\langle J| \int d^3x \fr{4({\bf E\cdot B})\phi}{M_{P\gamma}}  | J \rangle = \fr{\kappa}{M_{P\gamma}} 
\left(\fr{{\bf J}}{|J|}\cdot \nabla \phi\right),
\ee
where $\kappa$ is a dimensionless matrix element that can be calculated explicitly. 
For the ground state of the hydrogen atom, $\kappa$ is given by
\be
\kappa= \fr{8e\mu_B}{3a_0}=\fr{4\alpha^2}{3},
\label{hydrogen}
\ee
where $a_0$ and $\mu_B$ are Bohr radius and magneton, and $\alpha$ is the fine structure constant. 
This calculation take into account the magnetic field generated by the electron magnetic moment, and 
the electric field of the proton. If we consider both ${\bf E}$ and ${\bf B}$ created by the electron, 
we discover that the result has a logarithmic divergence in the ultraviolet regime that has the 
interpretation of $1/M_{Pe}$ being generated by $1/M_{P\gamma}$. Even with a modestly low choice of the 
cutoff, the coefficient is going to be on the order of $\alpha/\pi\sim O(10^{-3})$ and thus parametrically 
larger than (\ref{hydrogen}). 

What happens if instead of an atomic electron we consider a nucleus where the electric field 
is considerably stronger? To understand the scaling of the effect with $Z$, we consider a simplified
case of a single $s$-wave neutron above the closed nuclear shells with "uniform" distribution of its 
wave function inside the nucleus, 
which also has uniform charge distribution within a sphere of radius $R_N\simeq 1.2\,{\rm fm}\,(A)^{1/3}$. The
resulting $\kappa$ can be expressed in terms of the neutron magnetic moment,
\be
\kappa= \fr{8}{5}~\fr{2 \mu_n Z e}{R_N}= \fr{4}{5}~\fr{g_n Z \alpha}{m_pR_N} = 0.05-0.07~~~{\rm for}~~  
Z\sim 80,
\label{nucleus}
\ee
where the overall numerical coefficient follows from the approximation of the radial matrix element,  
$\langle r^2/(2R_N^2)-3/2\rangle_{r<R_N}=-6/5$. Although an overall numerical coefficient in 
estimate (\ref{nucleus}) cannot be taken very seriously, the parametric dependence on $Z$, $\mu_n$ and 
$R_N$ is certainly expected to hold for large nuclei. For mercury this effect is larger than the 
loop-induced admixture of the photon coupling into the nucleon coupling. Thus we can deduce 
the sensitivity 
of spin precession experiments to the 
pseudoscalar couplings to photons at 5\% level from the coupling to neutrons:
\be
|M_{P\gamma}M_S| \ga O (10^{-5})~M_{\rm Pl}^2~~~{\rm Ref.}~[29]
\ee 
One can see that the combined bounds from the clock comparison experiments are comparable to or
better than the product of separate bounds (\ref{Cassini}) and (\ref{limstars}). Unfortunately, these
bounds do not allow to probe the pseudoscalar coupling to fermions all the way to the 
"natural" scale $M_P \sim M_{\rm Pl}$.

\section{Conclusions}

Our paper considers the constraints on the combination of scalar and pseudoscalar 
couplings in the scalar-tensor theories of gravity. The strongest constraints 
come from the considerations of cosmological evolution of polarized light, and in the 
best case scenario of the maximal scalar coupling, consistent with 
constraints on Brans-Dicke theories, the sensitivity to the pseudoscalar coupling to photons can 
be as large as the Planck scale. However, the cosmological constraints are not sensitive to 
the derivative pseudoscalar couplings to fermions as they do not induce corresponding photon 
couplings even at the loop level. We also point out that for a wide range of pseudoscalar masses, 
one can avoid cosmological constraints due to the red-shifting of $\phi$-oscillations. 
Therefore, the laboratory constraints on spin precession 
from locally generated gradient of $\phi$ are complimentary
to cosmological bounds.
We revisited lab bounds to find that the most sensitive experiments are still few orders of 
magnitude below the sensitivity to Planck-scale-suppressed couplings. We also note that the 
local spin precession experiments provide sensitivity to the pseudoscalar coupling to 
photons, through the relatively large matrix element of $\phi {\bf B\cdot E}$ interaction inside 
atomic nuclei. 
As a separate remark, we have shown that nuclei of atoms used in the high-precision
clock comparison experiments have significant proton contribution to their spins. 
This allows to set separate constraints on pseudoscalar couplnigs to neutrons and 
protons, and improve the limit on Lorentz-violating axial-vector backgrounds in the 
proton sector. Further progress in experiments searching for a preferred Lorentz frame would also 
provide better sensitivity to the scalar-tensor theories extended by pseudoscalar couplings. 

{\bf Acknowledgements.} 
The authors would like to express their gratitude to W. Israel, who took an 
active part in the initial discussions that led to this project. 
M.P. would like to acknowledge useful conversations with M. Romalis. 
V.F. and S.L. thank Perimeter Institute for hospitality. 
M.P. would like to acknowledge the support of Gordon Godfrey fellowship at 
the UNSW. Research at the Perimeter Institute 
is supported in part by the Government
of Canada through NSERC and by the Province of Ontario through MEDT.


\begin{thebibliography}{99}

\bibitem{SCP} A.~G.~Riess {\it et al.}  [Supernova Search Team Collaboration],
Astron.\ J.\  {\bf 116} (1998) 1009
[arXiv:astro-ph/9805201];
S.~Perlmutter {\it et al.}  [Supernova Cosmology Project Collaboration],
Astrophys.\ J.\  {\bf 517} (1999) 565
[arXiv:astro-ph/9812133];
N.~A.~Bahcall, J.~P.~Ostriker, S.~Perlmutter and P.~J.~Steinhardt,
Science {\bf 284} (1999) 1481
[arXiv:astro-ph/9906463].

\bibitem{LSMG} See {\em e.g.} recent attempts in 
G.~R.~Dvali, G.~Gabadadze and M.~Porrati,
  Phys.\ Lett.\  B {\bf 485} (2000) 208; J.~D.~Bekenstein,
  Phys.\ Rev.\  D {\bf 70} (2004) 083509
  [Erratum-ibid.\  D {\bf 71} (2005) 069901];  N.~Arkani-Hamed, H.~C.~Cheng, M.~A.~Luty and S.~Mukohyama,
  JHEP {\bf 0405}, 074 (2004).
  
\bibitem{quint} B. Ratra, P.J.E. Peebles, Phys.\ Rev.\ D {\bf 37}, 3406 (1988);
Ap.\ J.\ Lett {\bf 325}, 117 (1988); C. Wetterich, Nucl.\ Phys.\ B {\bf 302},
668 (1988); R.R. Caldwell, R. Dave and P.J. Steinhardt, Phys.\ Rev.\ Lett {\bf
80}, 1582 (1998) [astro-ph/9708069].


\bibitem{Amendola}
L.~Amendola,
Phys.\ Rev.\ D {\bf 60}, 043501 (1999)
[arXiv:astro-ph/9904120];
L.~Amendola,
Phys.\ Rev.\ D {\bf 62}, 043511 (2000)
[arXiv:astro-ph/9908023].

\bibitem{Webb}
J.~K.~Webb {\it et al.},
Phys.\ Rev.\ Lett.\  {\bf 87}, 091301 (2001)
[arXiv:astro-ph/0012539];
M.~T.~Murphy, J.~K.~Webb and V.~V.~Flambaum,
Mon.\ Not.\ Roy.\ Astron.\ Soc.\  {\bf 345}, 609 (2003)
[arXiv:astro-ph/0306483].

\bi{Petitjean}
H.~Chand, R.~Srianand, P.~Petitjean and B.~Aracil,
Astron.\ Astrophys.\  {\bf 417}, 853 (2004)
[arXiv:astro-ph/0401094];
R.~Srianand, H.~Chand, P.~Petitjean and B.~Aracil,
Phys.\ Rev.\ Lett.\  {\bf 92}, 121302 (2004)
[arXiv:astro-ph/0402177].

\bi{Ralston}  B.~Nodland and J.~P.~Ralston,
  Phys.\ Rev.\ Lett.\  {\bf 78} (1997) 3043.


\bi{AntiRalston} S.~M.~Carroll and G.~B.~Field,
  Phys.\ Rev.\ Lett.\  {\bf 79} (1997) 2394;
J.~F.~C.~Wardle, R.~A.~Perley and M.~H.~Cohen,
  Phys.\ Rev.\ Lett.\  {\bf 79} (1997) 1801. 

\bi{ch_alpha} A representative (but not exhausitve) list of references 
on Bekenstein-type models includes: 
J.~D.~Bekenstein,
Phys.\ Rev.\ D {\bf 25} (1982) 1527;
T. Damour and A.M. Polyakov, Nucl.\ Phys.\ B {\bf 423}, 532 (1994);
M. Livio and M. Stiavelli, Ap. J. Lett. {\bf 507} (1998) L13; 
S.~J.~Landau and H.~Vucetich,
Astrophys.\ J.\  {\bf 570}, 463 (2002)
[arXiv:astro-ph/0005316];
H.~B.~Sandvik, J.~D.~Barrow and J.~Magueijo,
Phys.\ Rev.\ Lett.\  {\bf 88}, 031302 (2002); 
G.~R.~Dvali and M.~Zaldarriaga,
Phys.\ Rev.\ Lett.\  {\bf 88}, 091303 (2002);
T.~Damour, F.~Piazza and G.~Veneziano,
Phys.\ Rev.\ Lett.\  {\bf 89} (2002) 081601;
C. Wetterich, Phys.\ Lett.\ B {\bf 561}, 10 (2003); 
T.~Chiba and K.~Kohri,
Prog.\ Theor.\ Phys.\  {\bf 107} (2002) 631; C. Wetterich, Phys.\ Lett.\ B {\bf 561}, 10 (2003);  
L.~Anchordoqui and H.~Goldberg,
Phys.\ Rev.\ D {\bf 68}, 083513 (2003);
E.~J.~Copeland, N.~J.~Nunes and M.~Pospelov,
Phys.\ Rev.\ D {\bf 69}, 023501 (2004).

\bibitem{OP} K.~A.~Olive and M.~Pospelov,
Phys.\ Rev.\ D {\bf 65}, 085044 (2002).


\bi{Oklo} Nuclear physics derived constraints on changing couplings can be found in
A. I. Shlyakhter, Nature {\bf 264} (1976) 340; 
T.~Damour and F.~Dyson,
Nucl.\ Phys.\ B {\bf 480}, 37 (1996);
Y.~Fujii {\it et al.},
Nucl.\ Phys.\ B {\bf 573}, 377 (2000);
K.~A.~Olive, M.~Pospelov, Y.~Z.~Qian, A.~Coc, M.~Casse and E.~Vangioni-Flam,
Phys.\ Rev.\ D {\bf 66}, 045022 (2002);
K.~A.~Olive, M.~Pospelov, Y.~Z.~Qian, G.~Manhes, E.~Vangioni-Flam, A.~Coc and M.~Casse,
Phys.\ Rev.\ D {\bf 69}, 027701 (2004);
Y.~Fujii and A.~Iwamoto,
Phys.\ Rev.\ Lett.\  {\bf 91}, 261101 (2003).


\bi{Uzan} For recent reviews, see {\em e.g.} J.~P.~Uzan,
Rev.\ Mod.\ Phys.\  {\bf 75} (2003) 403; V.~V.~Flambaum,
  arXiv:0705.3704 [physics.atom-ph].



\bi{ph-sc}   A.~A.~Anselm,
  Yad.\ Fiz.\  {\bf 42} (1985) 1480; G.~Raffelt and L.~Stodolsky,
  Phys.\ Rev.\  D {\bf 37} (1988) 1237; C.~Deffayet and J.~P.~Uzan,
  Phys.\ Rev.\  D {\bf 62} (2000) 063507;
 C.~Csaki, N.~Kaloper and J.~Terning,
  Phys.\ Rev.\ Lett.\  {\bf 88}, 161302 (2002).
  
\bi{Opt_act}  J.~N.~Clarke, G.~Karl and P.~J.~S.~Watson,
  Can.\ J.\ Phys.\  {\bf 60} (1982) 1561; S.~M.~Carroll, G.~B.~Field and R.~Jackiw,
  Phys.\ Rev.\  D {\bf 41} (1990) 1231; A.~Lue, L.~M.~Wang and M.~Kamionkowski,
  Phys.\ Rev.\ Lett.\  {\bf 83} (1999) 1506;  A.~Kostelecky and M.~Mewes,
  arXiv:astro-ph/0702379;
  G.~C.~Liu, S.~Lee and K.~W.~Ng,
  Phys.\ Rev.\ Lett.\  {\bf 97} (2006) 161303.

\bi{Carroll} S.~M.~Carroll,
  Phys.\ Rev.\ Lett.\  {\bf 81}  (1998) 3067.
  
  
\bi{Opt_act_cmb}  
  B.~Feng, M.~Li, J.~Q.~Xia, X.~Chen and X.~Zhang,
  Phys.\ Rev.\ Lett.\  {\bf 96}, 221302 (2006);  
P.~Cabella, P.~Natoli and J.~Silk,
  arXiv:0705.0810 [astro-ph].


  

\bi{Will} C.~M.~Will,
  Living Rev.\ Rel.\  {\bf 9} (2006) 3.
  
\bi{MW} J.~E.~Moody and F.~Wilczek,
  Phys.\ Rev.\  D {\bf 30} (1984) 130.

\bi{BDD} T.~Banks, M.~Dine and M.~R.~Douglas,
  Phys.\ Rev.\ Lett.\  {\bf 88} (2002) 131301



\bibitem{Ellis} J.~R.~Ellis and M.~Karliner,
  arXiv:hep-ph/9510402.
  
  \bibitem{Cassini} B. Bertotti, L. Iess and P. Tortora, Nature {\bf 425} (2003) 374.


\bibitem{Raffelt} G.~G.~Raffelt,
  Ann.\ Rev.\ Nucl.\ Part.\ Sci.\  {\bf 49}, 163 (1999).

  
  \bibitem{WMAP-pol} A.~Kogut {\it et al.}  [WMAP Collaboration],
  Astrophys.\ J.\ Suppl.\  {\bf 148} (2003) 161
  [arXiv:astro-ph/0302213].

  
  \bibitem{Kronberg} P.P Kronberg, C.C. Dyer, amd H.-J. R\"oser, Astrophys. J. {\bf 472} (1996) 115. 


\bi{sg-old}I.~Y.~Kobzarev and L.~B.~Okun,
  Zh.\ Eksp.\ Teor.\ Fiz.\  {\bf 43} (1962) 1904
  [Sov.\ Phys.\ JETP {\bf 16} (1963) 1343]; J. Leitner abd S. Okubo, Phys.\ Rev.\ {\bf 136} (1964) 1542.
  
  \bi{sg-med} F.~W.~Hehl and W.-T.~Ni, Phys. Rev. {\bf D42} (1990) 2045;
  I.~B.~Khriplovich and A.~A.~Pomeransky,
  J.\ Exp.\ Theor.\ Phys.\  {\bf 86} (1998) 839
  [Zh.\ Eksp.\ Teor.\ Fiz.\  {\bf 113} (1998) 1537];
  A.~A.~Pomeransky, R.~A.~Senkov and I.~B.~Khriplovich,
  Phys.\ Usp.\  {\bf 43} (2000) 1055
  [Usp.\ Fiz.\ Nauk {\bf 43} (2000) 1129];  B.~Mashhoon,
  Lect.\ Notes Phys.\  {\bf 702} (2006) 112.

\bi{sg-new} A.~J.~Silenko and O.~V.~Teryaev,
  Phys.\ Rev.\  D {\bf 71}, 064016 (2005); A.~J.~Silenko and O.~V.~Teryaev,
  arXiv:gr-qc/0612103.

\bi{HgEDM} M. A. Rosenberry and T. E. Chupp, 
Phys.\ Rev.\ Lett.\  {\bf 86} (2001) 22; M.~V.~Romalis, W.~C.~Griffith and E.~N.~Fortson,
  Phys.\ Rev.\ Lett.\  {\bf 86} (2001) 2505; . 

\bi{Hg-Hg} B.~J.~Venema {\em et al.}, Phys. Rev. Lett. {\bf 68} (1992) 135.

\bi{gs} D. J. Wineland {\em et al.}, Phys. Rev. Lett. {\bf 67} (1991) 1735;
D. J. Wineland  and N. F. Ramsey, Phys. Rev. {\bf A5} (1972) 821. 

\bi{Berglund} C. J. Berglund {\em et al.}, Phys. Rev. Lett. {\bf 75} (1995) 1879.

\bi{Bear}  D.~Bear, R.~E.~Stoner, R.~L.~Walsworth, V.~A.~Kostelecky and C.~D.~Lane,
  Phys.\ Rev.\ Lett.\  {\bf 85} (2000) 5038
  [Erratum-ibid.\  {\bf 89} (2002) 209902].


\bi{Heckel} B.~R.~Heckel, C.~E.~Cramer, T.~S.~Cook, E.~G.~Adelberger, S.~Schlamminger and U.~Schmidt,
  Phys.\ Rev.\ Lett.\  {\bf 97} (2006) 021603. 

\bi{Kost} D.~Colladay and V.~A.~Kostelecky,
  Phys.\ Rev.\  D {\bf 58} (1998) 116002.

\bi{KostLane}  V.~A.~Kostelecky and C.~D.~Lane,
  Phys.\ Rev.\  D {\bf 60}, 116010 (1999).



\bi{Walsworth} 
  D.~F.~Phillips, M.~A.~Humphrey, E.~M.~Mattison, R.~E.~Stoner, R.~F.~C.~Vessot and R.~L.~Walsworth,
  Phys.\ Rev.\  D {\bf 63}, 111101 (2001)















\end{thebibliography}
\end{document}